\documentclass[twocolumn,secnumarabic,amssymb, nobibnotes, aps, prd]{revtex4-1}
\usepackage{graphicx}
\usepackage{bm,color}
\usepackage{mathrsfs}
\usepackage{verbatim}
\usepackage{braket}
\usepackage{subfigure}

\graphicspath{{figures/}}
\begin{document}

\title{Spin fluctuations in the light-induced high spin state of Cobalt valence tautomers}
\author{ F. Caracciolo$^{1}$, M. Mannini$^{2}$, G. Poneti$^{3}$, M. Pregelj$^{4}$, N. Jan\v{s}a$^{4}$,  D. Ar\v{c}on$^{4}$$^{,5}$,  P. Carretta$^{1}$}
\affiliation{$^{1}$Department of Physics, University of Pavia,
Pavia, I-27100 Italy} \affiliation{$^{2}$Department of Chemistry,
University of Florence $\&$ INSTM RU Florence, Sesto Fiorentino,
I-50019 Italy} \affiliation{$^{3}$Instituto de Qu\'imica,
Universidade Federal do Rio de Janeiro,  Rio de Janeiro,
21941-909 Brazil} \affiliation{$^{4}$Jo$\check{z}$ef Stefan Institute,
Ljubljana, Slovenia}\affiliation{$^{,5}$Faculty of Mathematics and Physics, University of Ljubljana,Ljubljana SI-1000, Slovenia}

\begin{abstract}

We present a study of the static magnetic properties and spin
dynamics in Cobalt valence tautomers (VT), molecules where a
low-spin (LS) to high-spin (HS) crossover driven by an
intramolecular electron transfer can be controlled by the
temperature, by the external pressure or by light irradiation. In
the investigated complex, a LS-Co(III) ion bound to a dinegative
organic ligand can be reversibly converted into the HS-Co(II)
bound to a mononegative one. By combining magnetization
measurements  with Nuclear Magnetic Resonance (NMR) and Muon Spin
Relaxation ($\mu$SR), we have investigated the static magnetic
properties and the spin dynamics as a function of the temperature.
Moreover, the effect of the external pressure as well as of the
infrared light irradiation have been explored through magnetometry
and NMR measurements to determine the spin dynamics of the HS
state. The photoinduced HS state, which can have a lifetime of
several hours below 30 K, is characterized by spin dynamics in the
MHz range, which persist at least down to 10 K. The
application of an external pressure causes a progressive increase
of the LS-HS crossover, which reaches room temperature for
pressures around 10 kbar.

\end{abstract}

\maketitle

\section{Introduction}

In the trend towards miniaturization of magnetic data storage
media, the ultimate size reduction of the information units is
limited by the superparamagnetism of nanosized magnetic media as
well as by the complexity of achieving controlled electrical
contacts at the nanoscale with a low dissipation
\cite{White2000,Richter2009,Moser2002}. An appealing solution to
both challenges would be a combination of functional switchable
molecules as bits and the use of light irradiation as read/write
protocols \cite{Andreasson2015,Pischel2010,Dvornikov2009}. Single
molecules have already shown to mimic electronic device
architectures, suggesting the viability of this approach
\cite{Song2011,Bogani2008,Mannini2009,Weibel2007}. Among them,
iron(II)-based spin crossover complexes are a paradigmatic example
of bistable materials where the magnetic state can be changed from
a low-spin (LS) diamagnetic configuration into a high-spin (HS)
paramagnetic one through a variety of external stimuli, including
light irradiation, temperature variation and pressure application
\cite{Halcrow2013,Decurtins1984,Decurtins1985,Gutlich2005}.

\begin{figure}[h]
\centering \subfigure[]{\includegraphics[width=9 cm, height=7
cm,trim={0 0 0 0}]{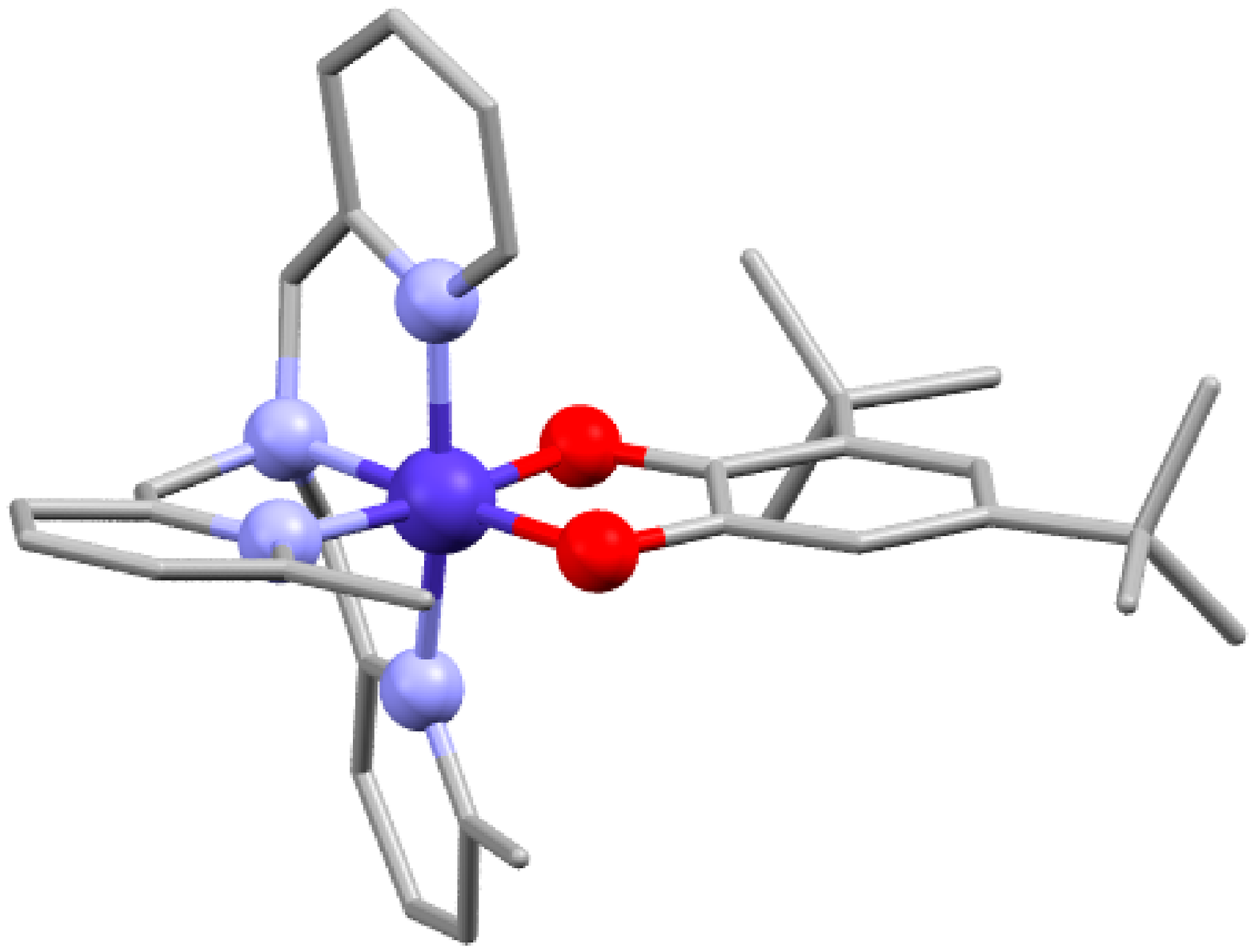}}
\subfigure[]{\includegraphics[width=11 cm, height=8 cm,trim={0
0 0 0}]{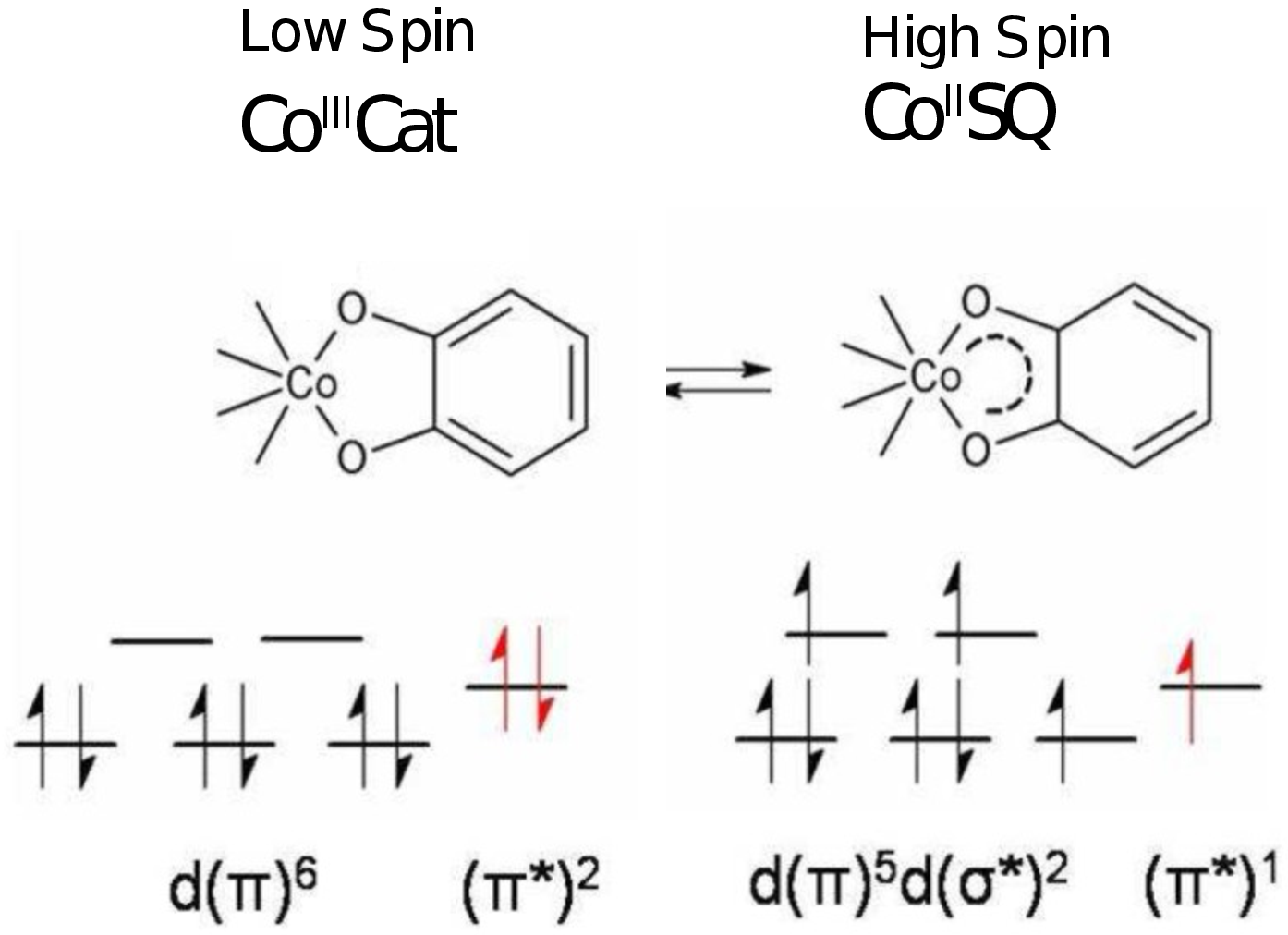}} \caption{a) Molecular structure of
Co(Me$_{2}$tpa)(DBCat)]PF$_{6}$ (Me$_{2}$tpa = bis(6-methyl-(2-pyridylmethyl))(2-pyridylmethyl)amine, DBCat = 3,5-di-tert-butylcatecholato). Color
code: Cobalt (grey ball), oxygen (red ball), nitrogen (violet ball), carbon
(grey stick). Hydrogen atoms are omitted for clarity. b) The two
electronic configurations involved in the VT interconversion
yielding the LS diamagnetic and the HS paramagnetic state.}
\label{Fig:structure}
\end{figure}

Another class of materials characterized by the possibility of
reversibly switching among different electronic states is the one
of Valence Tautomers (VT)
\cite{Tezgerevska2014,Gutlich1997,Pierpont2001,Adams1993,Carbonera2004,Tao2006,Caneschi1998,Dei2010}.
VT are constituted by a metal ion bound to a redox-active ligand: in
these systems the two states can be interconverted by an
intramolecular electron transfer between the metal ion and the
ligand, followed by a change of the spin state of the ion. In the
most investigated family of VTs, the two main building blocks are
a cobalt ion and an organic ligand belonging to the dioxolene
family (Fig. \ref{Fig:structure}a). Two different electronic states
can be realized (Fig. \ref{Fig:structure}b): LS-Co(III) coupled to the ligand in its
binegative diamagnetic state (Co(III)-catecholato state, with two
electrons in the $\pi$ organic ligand molecular orbital) and
HS-Co(II) bound to the ligand in its oxidized mononegative radical
form (Co(II)-semiquinonato state, with only one electron in the $\pi$ orbital
of the ligand) \cite{Dapporto2008} (we will refer to the low spin and high spin magnetic states with LS and HS, respectively, for the sake of simplicity). The interconversion between
the two states can be driven by the temperature. The technological interest for these systems originates from the
possibility to control the spin state via thermal, structural, optical and
magnetic perturbations \cite{Poneti2010}, suggesting the use of
these materials in the next generation of molecular spintronic
devices \cite{Droghetti2011}. Moreover, it is possible to
nanostructure them on a solid substrate via wet chemistry
strategies, while keeping switchability at the nanoscale
\cite{Poneti2015,Vazquez2016}. Furthermore, VT are, together with
spin-crossover systems, good candidates for pressure sensors
\cite{Roux1996,Caneschi2001,Li2008b,Gutlich2004}, since the VT
transition can be tuned by pressure and be revealed by a change in
the light absorption spectrum of the material.

Remarkably, at low temperature cobalt-dioxolene electronic state can be
switched with light irradiation \cite{Sato2007,Dapporto2008}. In
order to exploit this feature to implement them as logic units, it
is necessary to investigate both the lifetime of the photoinduced
HS state as well as its spin dynamics. To determine the
former, we have used a SQUID magnetometer coupled to an IR laser
diode, while for the latter we have investigated the spin dynamics
by means of magnetic resonance techniques. In particular, it is of major importance to determine how the spin
fluctuations in VT evolve with temperature when the system is
driven to the HS state by light irradiation, by using local-probe
techniques as nuclear magnetic resonance (NMR) and muon spin
relaxation ($\mu$SR), since these fluctuations determine how
long the information will be preserved before it gets lost.
Moreover these techniques can be used to detect the effect of an
external pressure on the VT state \cite{Sato2002,Adams1995,Sato2007,Dapporto2008,Poneti2015,Cui2004,Dai2013}.

We start by discussing the results of the SQUID experiments, which
allowed to derive the temperature dependence of the magnetization,
its build-up times under light irradiation and the effect of the
external pressure on the LS-HS crossover temperature. Then we show
how $^{1}$H NMR and $\mu$SR allow us to investigate the low
temperature dynamics of the HS molecules and the microscopic
effects of the light and external pressure. Finally, $^{59}$Co NMR
is exploited to deeply investigate the spin fluctuations of the
light induced HS molecules at low temperatures.

\section{Experimental methods and results}

\subsection{SQUID}

\begin{figure}[h]
\centering
\includegraphics[width=9 cm, height=6 cm]{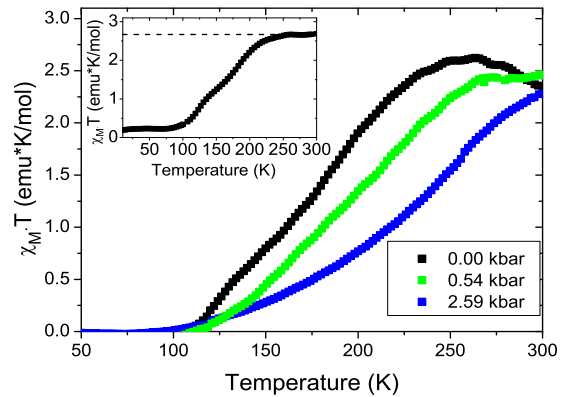}
\caption{Temperature dependence of $\chi_{M}\cdot T$ for different values
of the applied pressure, for a 0.1 T magnetic field. The ambient pressure measurement (P = 0.00 kbar) has been performed inside the pressure cell and it is affected by its background contribution, yielding a slight different shape from the inset of Fig. \ref{Fig:Mag_P}. The inset shows the temperature dependence of the VT magnetic susceptibility multiplied by temperature, for a 0.1 T magnetic field \cite{Dapporto2008}. The horizontal dashed line refers to the high temperature plateau.}
\label{Fig:Mag_P}
\end{figure}

The synthesis of the powder sample was carried out following a
previously published procedure \cite{Dapporto2008}. The
temperature dependence of the magnetic susceptibility $\chi_{M}$
was measured with an MPMSXL7 SQUID magnetometer, by cooling (4
K/min) in 0.1 T (inset of Fig. \ref{Fig:Mag_P}). A clear reduction
of $\chi_{M}\cdot T$ is observed, with a broad crossover from the
HS to the LS state extending from 200 to 100 K and
centered around 160 K \cite{Dapporto2008}. This crossover most
likely reflects the presence of two overlapping transitions, which
are associated with crystallographically inequivalent molecules in
the lattice  \cite{Dapporto2008}.
The temperature dependence of $\chi_{M}\cdot T$ was also
investigated at different external pressures, by using a CuBe
pressure cell (EasyLab, Mcell 10). Even if the background
contribution from the cell modifies the shape of $\chi_M(T)$
curves, upon increasing the pressure (measured with a Sn
manometer) there is a clear progressive increase in the LS-HS
crossover temperature (Fig. \ref{Fig:Mag_P}).

\begin{figure}[h]
\centering
\includegraphics[width=9 cm, height=6 cm]{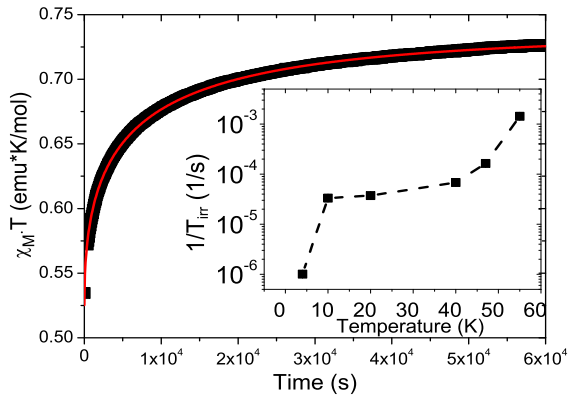}
\caption{$\chi_{M}\cdot T$, at $T= 47$ K for a 0.1 T magnetic
field, as a function of the time elapsed after the infrared light
has been turned on. The solid line is the best fit according to
Eq.\ref{Mag_irr}. The inset shows the temperature dependence of
the build-up rate in a 0.1 T magnetic field. The line is a guide
to the eye.} \label{Fig:mag_grow}
\end{figure}

In order to irradiate the sample an optical fiber was placed along
the SQUID stick and coupled with an external infrared Laser diode
(Thorlabs M940F1), with a central wavelength of 940 nm and a
measured output power around 6 mW at the end of the fiber. Under
light irradiation, in the 4 - 55 K temperature range,
$\chi_{M}\cdot T$ was found to increase with time (Fig.
\ref{Fig:mag_grow}). The susceptibility build-up can nicely be fit
with a stretched exponential recovery law plus a constant term
$\chi _{1}$ corresponding to the susceptibility value prior to
irradiation:
\begin{equation}
\chi _{M}(t) =\chi _{0}\cdot \biggl(1- e^{-{({t}/{T_{irr}}})^{\alpha_{M}}}
\biggr) + \chi _{1} \label{Mag_irr}.
\end{equation}
$\chi _{0}$ is the light enhanced magnetic
susceptibility, $T_{irr}$ the build-up time and $\alpha_{M}$ a
stretching exponent with values ranging from 0.50, at T = 4 K,
to 0.70, at T = 55 K.
The build-up time increases at low temperatures and
reaches several hours at 10 K.
After switching off the light irradiation the susceptibility decays
back to its thermal equilibrium value with an initial stretched
exponential decay \cite{Beni2006,Li2008}, with a different decay
time, reaching about $2 \cdot 10^{4}$ s at 20 K
(see Supplementary Material).



\subsection{Nuclear Magnetic Resonance}

\begin{figure}[h]
\centering
\includegraphics[width=9 cm, height=6 cm]{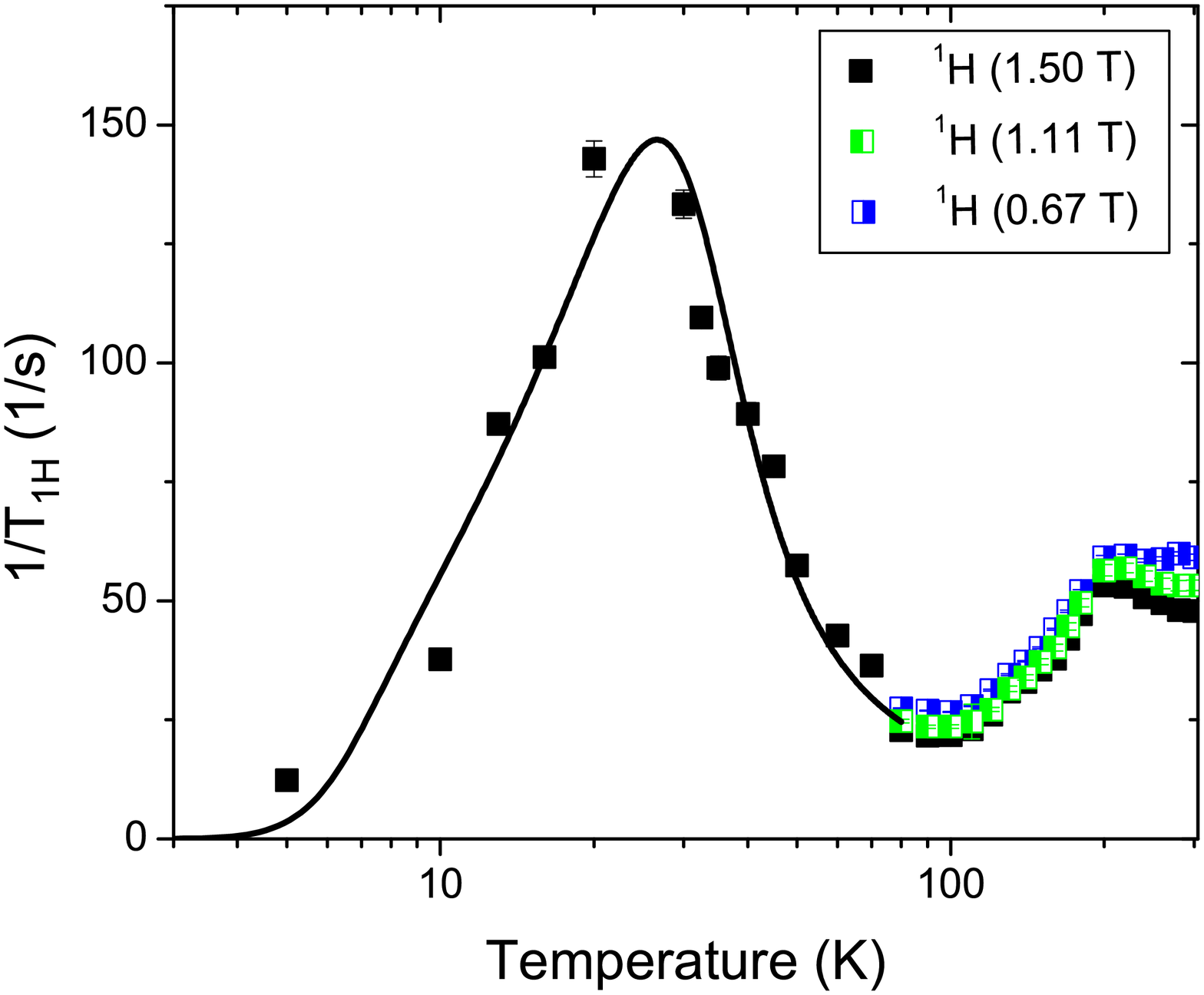}
\caption{Temperature dependence of $^1$H nuclear spin-lattice
relaxation rate with LEDs OFF at: 1.50 T (full black squares),
1.10 T (half left green squares) and 0.67 T (half right blue
squares). Fit according to Eq. \ref{BPP}. Fit parameters are
reported in Table \ref{Tbl:BPPCo}.} \label{Fig:T11H}
\end{figure}

\begin{figure}[h]
\centering
\includegraphics[width=9 cm, height=6 cm]{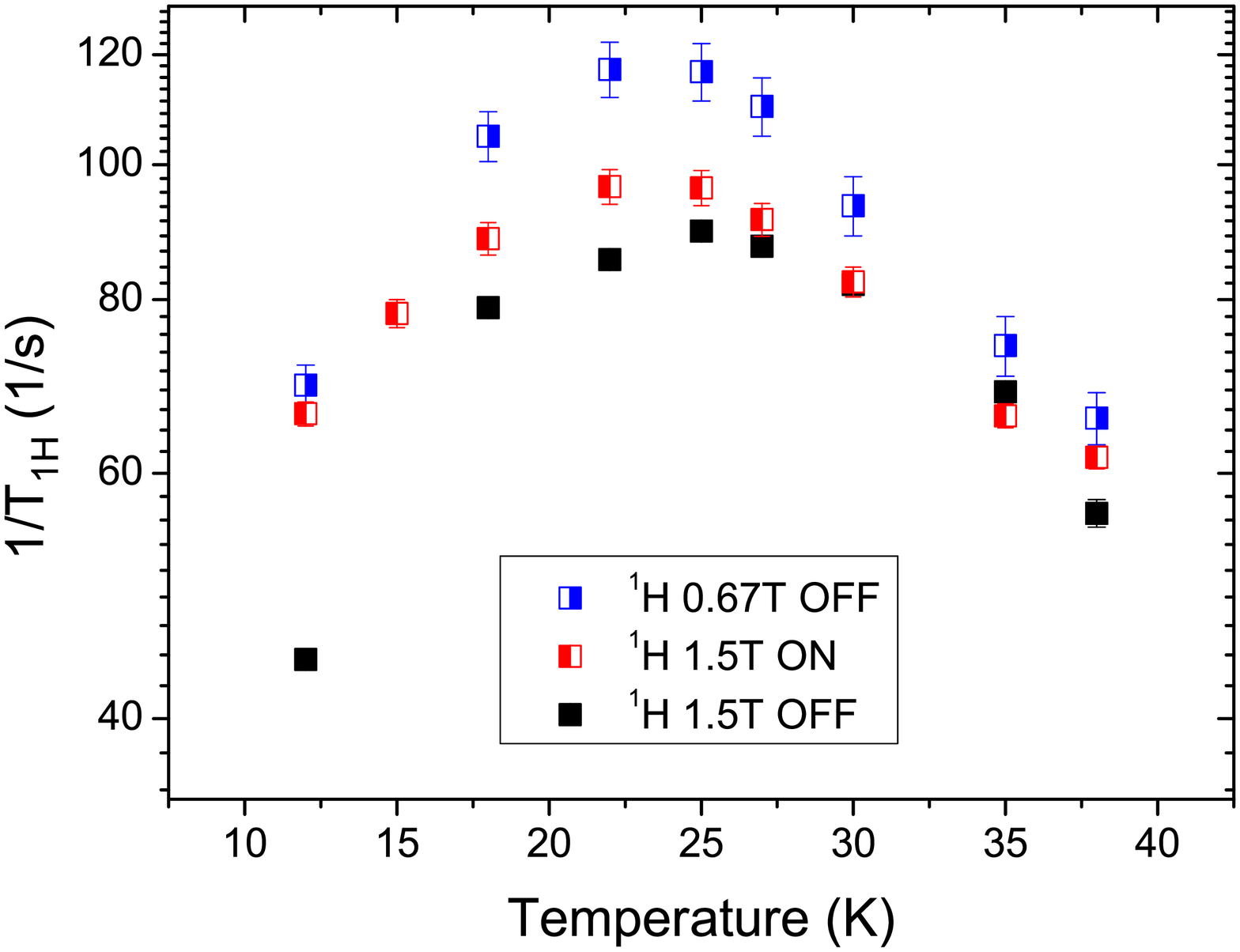}
\caption{Temperature dependence of $^1$H nuclear spin-lattice relaxation rate for different external magnetic fields:
1.50 T and LED OFF (full black squares), 1.50 T and LED ON (half left red squares), 0.67 T and LED OFF (half right blue squares). This measurements have been performed with a cooling rate of 6 K/min.}
\label{Fig:peakfield}
\end{figure}

$^1$H (nuclear spin $I=1/2$) NMR experiments have been performed
in 1.5, 1.1 and a 0.67 T in the 5 - 300 K
temperature range. 
 The $^1$H NMR spectrum was derived from the
Fourier transform of half of the echo signal after a solid echo
pulse sequence. The spectrum is characterized by a Gaussian shape
with a weakly temperature dependent linewidth, increasing from 20
to 30 kHz by decreasing the temperature down to 10 K. $^1$H
nuclear spin-lattice relaxation rate $1/T_{1H}$ was derived from
the recovery of nuclear magnetization $m(t)$ after a saturation
recovery pulse sequence. The recovery law was a stretched
exponential:
\begin{equation}
m(t)=m_{0}\cdot \biggl[1-e^{-({t}/{T_{1H}})^{\alpha_{H}}}\biggr]
\,\, , \label{rec}
\end{equation}
where $m_{0}$ is the $^1$H thermal equilibrium magnetization and
$\alpha_{H}$ a stretching exponent, ranging between 0.70 and 0.95,
which accounts for a distribution of relaxation times due to the
presence of inequivalent $^1$H sites.
The temperature dependence of $1/T_{1H}$ is shown in Fig.
\ref{Fig:T11H}, for different values of the external magnetic
field, for a cooling rate of 1 K/min. One notices a very slight
decrease of the relaxation rate around room temperature by
increasing the magnetic field
strength (see Fig \ref{Fig:T11H}). 
The temperature dependence of $1/T_{1H}$ is characterized by a
high temperature plateau, by a drop on cooling across the HS-LS
crossover and then by a marked peak centered around 20 K. This
peak was investigated also with a faster cooling rate (6 K/min) at
two different magnetic fields, as shown in Fig.
\ref{Fig:peakfield}. By comparing the results shown in Figs. 4 and
5 for $H=1.5$ T, one can notice that the peak in $1/T_1$ strongly
depends on the cooling rate. Moreover, by increasing the magnetic
field the peak intensity decreases.

Irradiation was achieved by using two 910 nm LEDs (Thorlabs
LED910E), with an output power of 8 mW, placed close to the
sample. The LEDs were switched on at room temperature, then the
sample was cooled down to 15 K, kept at this temperature for
several hours and then the measurements were carried out upon
increasing the temperature, with LEDs switched on. One
observes a slight increase of $1/T_{1H}$ low $T$ peak under light
irradiation (Fig. \ref{Fig:peakfield}).

\begin{figure}[h]
\centering
\includegraphics[width=9 cm, height=6 cm]{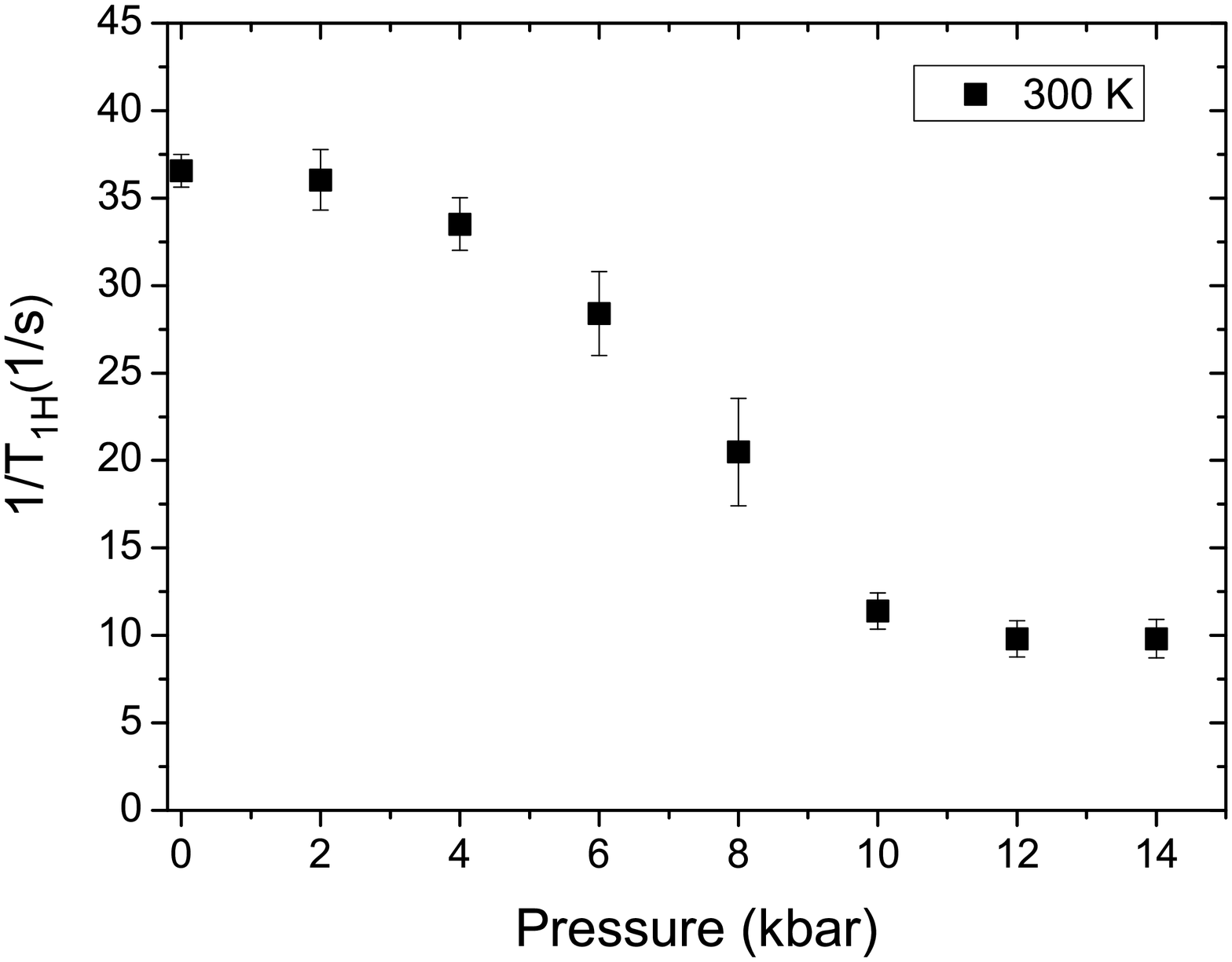}
\caption{$^1$H spin-lattice relaxation rate pressure dependence in
a 4.69 T magnetic field at 300 K.} \label{Fig:1H_P}
\end{figure}

To explore the effect of pressure on the LS-HS conversion, high
pressure $^1$H NMR experiments have been performed in a 4.69 T
magnetic field at 300 K.  The applied pressure ranged from 0 to 14
kbar, and was progressively increased in 2 kbar steps. The proton
spin-lattice relaxation rate pressure dependence is shown in Fig.
\ref{Fig:1H_P}.

\begin{figure}[h]
\centering
\includegraphics[width=9 cm, height=6 cm]{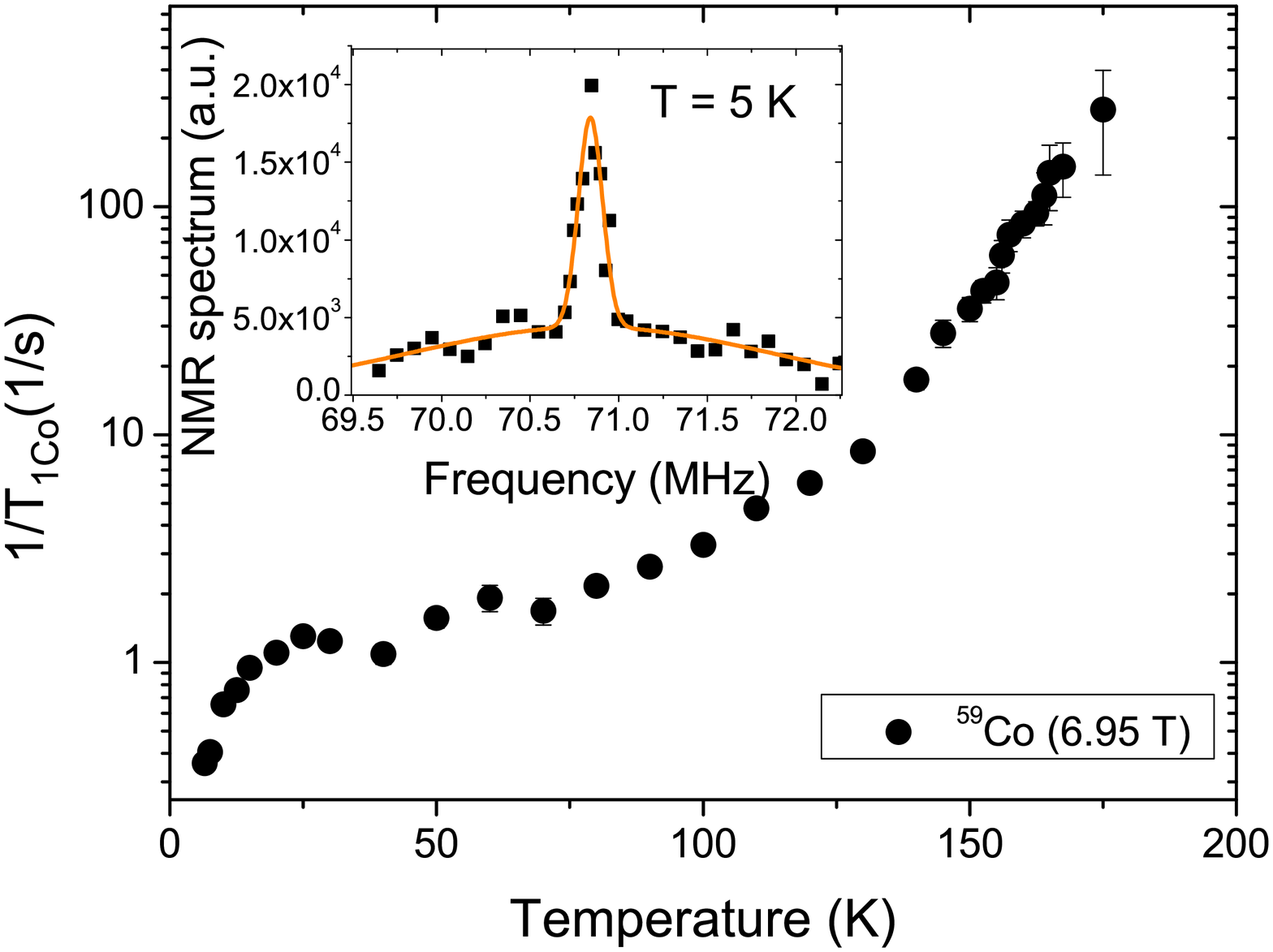}
\caption{Temperature dependence of $^{59}$Co nuclear spin-lattice
relaxation rate at 6.95 T. The inset graph shows $^{59}$Co
spectrum at 5 K at the same magnetic field (the line is a guide to the eye). 
}
\label{Fig:T1CoSpectrum}
\end{figure}

$^{59}$Co ($I=7/2$) NMR measurements have been performed in a 6.95
T magnetic field. $^{59}$Co NMR powder spectrum, broadened by the
quadrupolar coupling, was derived by recording the amplitude of
the echo obtained after a Hahn echo saturation recovery sequence upon varying
the irradiation frequency. The spectrum (see the inset to
Fig.\ref{Fig:T1CoSpectrum}, for T = 5 K) is characterized by a central peak,
associated with the $m_I= +1/2\leftrightarrow -1/2$ central
transition, and by featureless less intense shoulders associated
with the satellite transitions, which extend over several MHz.

\begin{figure}[h]
\centering
\includegraphics[width=9 cm, height=6 cm]{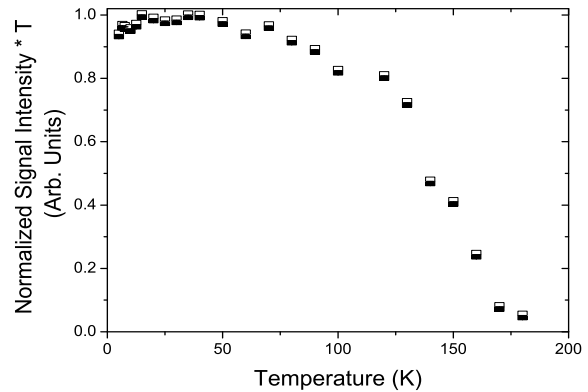}
\caption{Temperature dependence of $^{59}$Co NMR signal intensity, multiplied by temperature, in a 6.95 T magnetic field.}
\label{Fig:CoSignal}
\end{figure}

The spin-lattice relaxation rate $1/T_{1Co}$ has been measured irradiating the central transition. The intensity
of this line is observed to significantly decrease upon increasing
the temperature and the signal could be detected only up to about
180 K (Fig. \ref{Fig:CoSignal}). At higher temperature, where almost all VT are in the HS
state the fast nuclear relaxation prevents the observation of the
$^{59}$Co NMR signal. In other terms, the signal being detected is
the one arising from nuclei of VT molecules in the LS state.
$1/T_{1Co}$ was derived by fitting the recovery of the nuclear
magnetization after a saturation recovery pulse sequence with the
recovery law for the central NMR line of $I=7/2$ nuclei
\cite{Suter1998}.
The temperature dependence of $1/T_{1Co}$ is shown in
Fig.\ref{Fig:T1CoSpectrum}.

Fig. \ref{Fig:CoSignal} shows the
decrease of the product of the normalized signal intensity with the
temperature.

\begin{figure}[h]
\centering
\includegraphics[width=9 cm, height=6 cm]{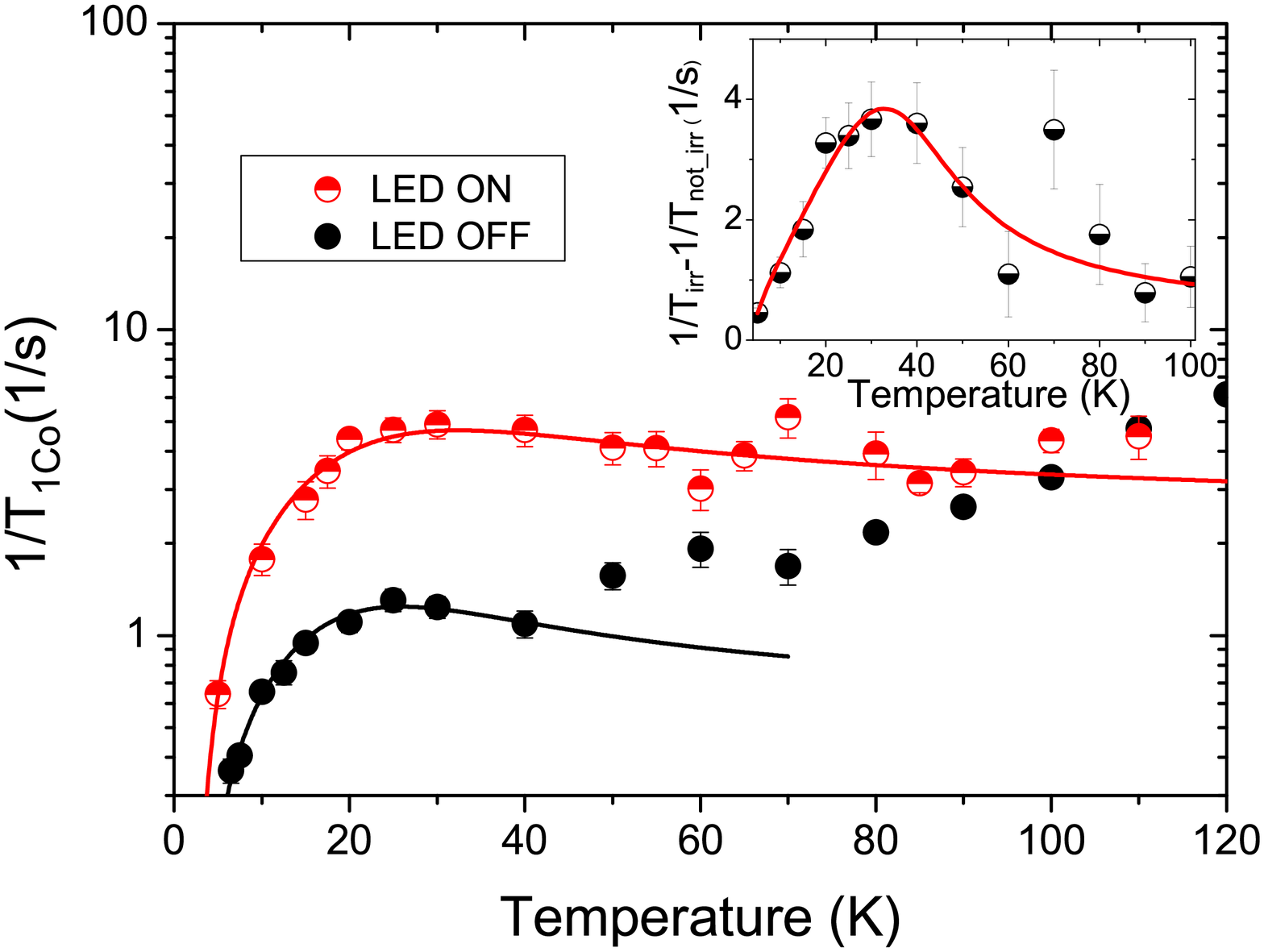}
\caption{Temperature dependence of $^{59}$Co nuclear spin-lattice
relaxation rate at 6.95 T: LED OFF (full black circles), LED ON
(half up red circles). Fit according to Eq. \ref{BPP}. Fit
parameters are reported in Table \ref{Tbl:BPPCo}. The inset shows the temperature dependence of the difference between $^{59}$Co nuclear spin-lattice relaxation rate in the presence and in the absence of infrared light irradiation. Fit according to Eq. \ref{BPP}. Fit parameters are reported in Table \ref{Tbl:BPPCo}.} \label{Fig:T1Co}
\end{figure}

As in the case of $^1$H NMR experiments, the sample was irradiated
with two LEDs (Thorlabs LED910E) mounted close to the sample and a
cooling rate of 6 K/min was used during the $^{59}$Co NMR
measurements, and the same heating sequence used for $^1$H NMR
measurements was followed (except that the sample was cooled down
to 5 K instead of 15 K). The light irradiation
is found to cause a significant increase in $1/T_{1Co}$
(Fig.\ref{Fig:T1Co}) below 100 K.

\subsection{Muon Spin Relaxation}

\begin{figure}[h]
\centering
\includegraphics[width=9 cm, height=6 cm]{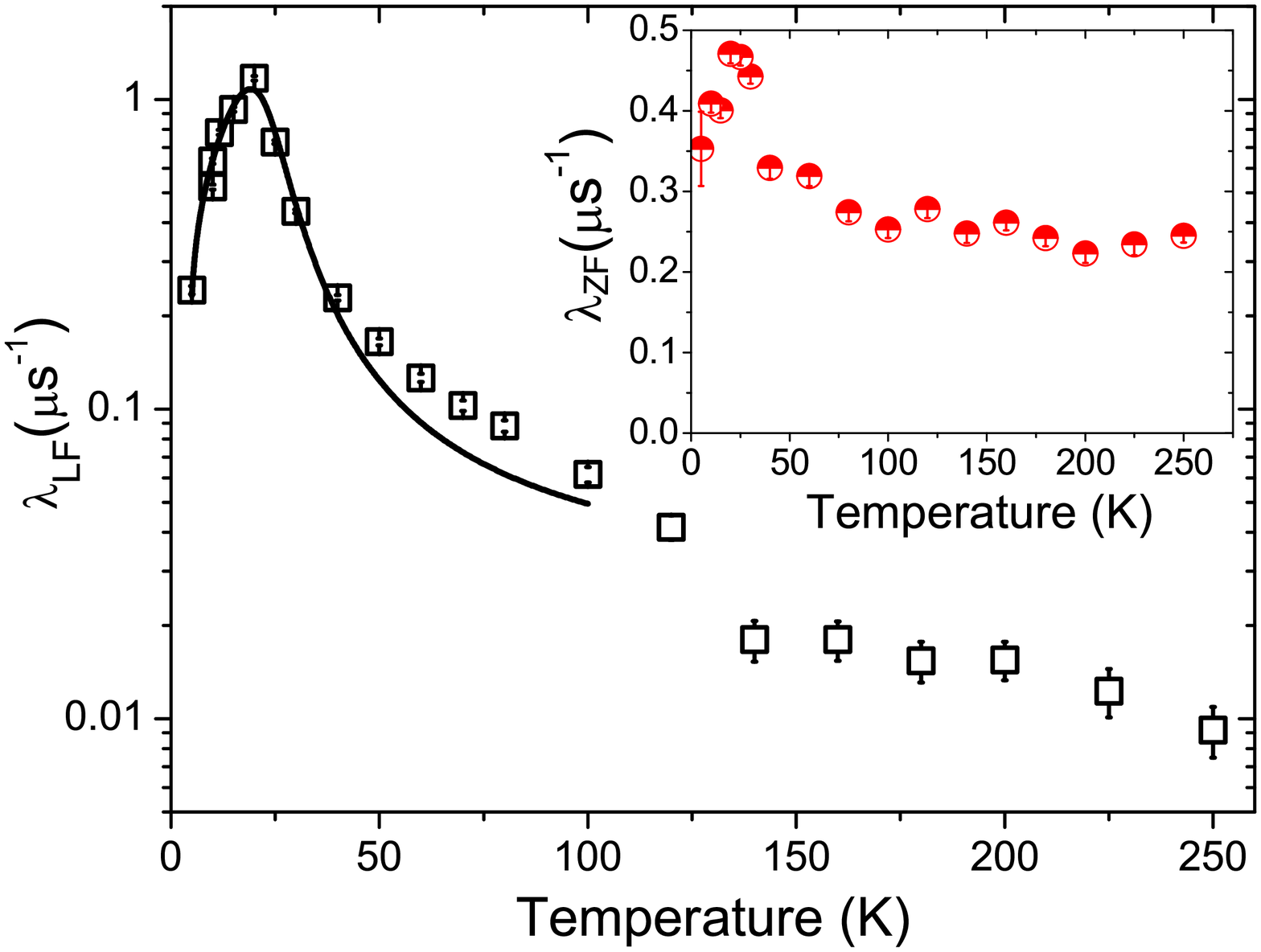}
\caption{Temperature dependence of the muon spin relaxation rate
for a 200 Gauss longitudinal magnetic field. Lines are fit
according to Eq. \ref{BPP}, with parameters: $\delta h_{\bot}^{2}$
= (3.2 $\pm$ 0.1) $\cdot 10^{-10}$ T$^{2}$, $\tau$ = (1.5 $\pm$
0.9)$\cdot 10^{-10}$ s, E$_A$ = 77 $\pm$ 5 K, $\Delta$ = 54 $\pm$
4 K. The inset shows the temperature dependence of the muon spin
relaxation rate in ZF experiments.} \label{Fig:lambdazf}
\end{figure}

$\mu$SR experiments have been carried out at ISIS facility on HIFI
beam line. The measurements have been performed both in Zero Field
(ZF) and in a 200 Gauss Longitudinal magnetic field (LF). The
decay of the muon asymmetry in ZF, after background corrections,
has been fit with a simple exponential decay
\begin{equation}
A(t)= A_1 e^{-\lambda_{ZF} t} \, , \label{relZF}
\end{equation}
where $\lambda_{ZF}$ is the muon's ZF decay rate. It is noticed that $A_1$ in zero field corresponds to about half of the total initial asymmetry, likely due to the presence of a muonium fraction \cite{Patterson1988}. The application of the longitudinal field causes both an increase in the initial asymmetry and a slowing down of the relaxation at high temperatures (see Supplementary Material). Muon's asymmetry in LF runs can be reproduced
with a stretched exponential decay:
\begin{equation}
A(t)= A_0 e^{-(\lambda_{LF} t)^{\beta_{LF}}}, \label{relLF}
\end{equation}
where $\lambda_{LF}$ is the longitudinal relaxation rate and
$\beta_{LF}$ a stretching exponent, with values ranging from about
0.35 at low temperatures to 0.60 at 20 - 25 K and to 0.50 at higher
temperature. As it can be noticed from Fig. \ref{Fig:lambdazf},
muon relaxation is faster at low temperature and both in ZF and LF
experiments a peak around 20 K is detected.

\section{Discussion}

The temperature dependence of
$\chi_M\cdot T$ (inset of Fig.\ref{Fig:Mag_P}) clearly shows that
all molecules are paramagnetic at high temperature and start
becoming diamagnetic upon decreasing the temperature below 200 K,
as expected for this compound.\cite{Poneti2010,Dapporto2008} The
high temperature plateau ($\chi_{M}\cdot T =$ 2.67 emu
$\cdot$K/mol, corresponding to 3.35 $\times$ 10$^{-5}$
m$^3$$\cdot$K/mol, there is a $4\pi\times 10^{-6}$ conversion
factor between CGS and SI units) is expected for an ensemble of
non interacting paramagnetic molecules:
\begin{equation}
\chi_{M}\cdot T = \frac{g^{2}\mu_{B}^{2}S(S+1)N_{A}}{3K_{B}},
\label{curie}
\end{equation}
where $N_{A}$ is the Avogadro number and $g\simeq 2$ the Land\'e
factor. The sum of the contribution from the uncoupled radical
(S=${1}/{2}$, $\chi_{M}\cdot T = 0.375$ emu $\cdot$K/mol) and from
the high spin Co(II) ($\chi_{M}\cdot T = 2.4-2.6$ emu $\cdot$K/mol
that is the sum of a S=3/2 contribution (see Fig. 1) with
$\chi_{M}\cdot T = 1.875$ emu $\cdot$K/mol and of an orbital term)
yields a value around 2.9 emu $\cdot$K/mol \cite{Dapporto2008},
close to the experimental one. 
On the other hand, a ferromagnetic coupling between the HS-Co(II) ion and the radical would give rise to a local S = 2 spin, which would yield a $\chi_{M}\cdot T$ value of about 3.2 - 3.4 emu $\cdot$K/mol, which is above the detected one.
Notice that an antiferromagnetic coupling, would yield an S=1 state and a value
for $\chi_{M}\cdot T$ = 1 emu $\cdot$K/mol, quite different from
the experimental. Hence, the configuration with a local S=3/2
Co(II) ion decoupled from the S=1/2 radical seems to provide the
best description for the high temperature HS state
\cite{Dapporto2008}.

The fraction $HS_{LT}$ of molecules which remain in the HS state
even at low temperatures is given by:
\begin{equation}
HS_{LT} = \frac{\chi_{M}-\chi_{LS}}{\chi_{pure}-\chi_{LS}},
\label{remaining}
\end{equation}
where $\chi_{LS}$ is the magnetic susceptibility of the LS state
and $\chi_{pure}$ is the magnetic susceptibility of an
isostructural compound, not displaying the VT transition, which
remains in the HS state over the whole temperature range
\cite{Beni2008}. It is found that about 3 $\%$ of the molecules
remain in the HS state at 5 K, a value close to that found in
the same compound
\cite{Tayagaki2005}.

Irradiation at cryogenic temperatures with infrared light promotes
molecules from the LS to the HS state and enhances the
magnetization (Fig. \ref{Fig:mag_grow})
\cite{Dapporto2008,Neuwahl2002,Gentili2005,Schmidt2010,Beni2007}.
The build-up times can reach several hours \cite{Schmidt2010},
making the whole process quite slow. The efficiency of the
conversion, namely the fraction of molecules converted in the HS
state, can be estimated as:
\begin{equation}
\epsilon = \frac{\chi_{O}-\chi_{M}}{\chi_{pure}-\chi_{M}}\,\,\,\,
. \label{efficiency}
\end{equation}

\begin{figure}[h]
\centering
\includegraphics[width=9 cm, height=6 cm]{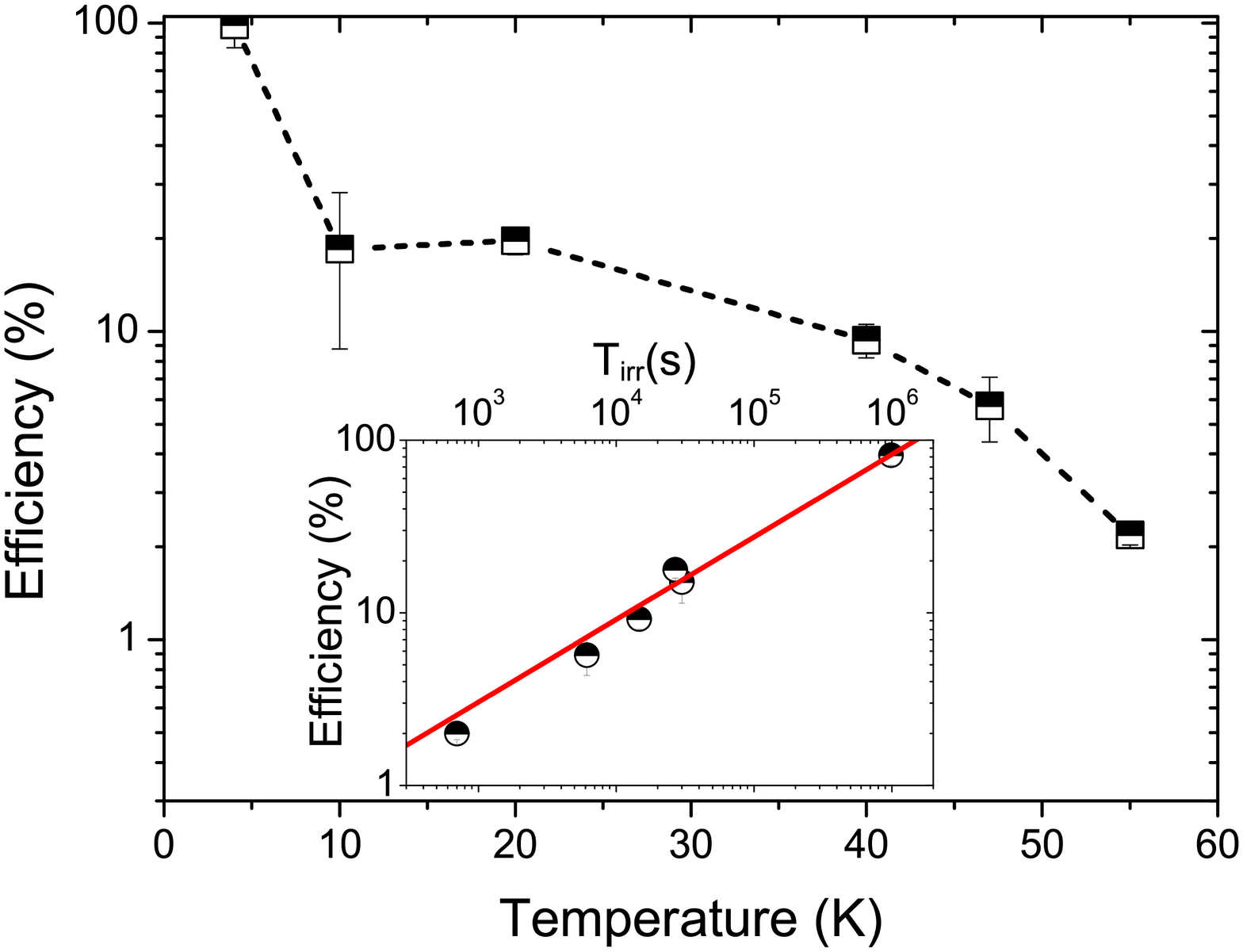}
\caption{Temperature dependence of IR irradiation efficiency, according
to Eq. \ref{efficiency}, as derived from SQUID experiments. The
line is a guide to the eye. The inset shows the irradiation
efficiency over build-up times, with temperature as an implicit
parameter. The line is a power-law fit with coefficient a= 0.11
$\pm$ 0.03 and exponent b= 0.48 $\pm$ 0.02.} \label{Fig:eff}
\end{figure}

As the temperature
is decreased the light conversion efficiency improves, due to the
increased HS lifetime (Fig. \ref{Fig:eff}). In other terms, as the
temperature increases, thermal relaxation, driving the VT back to
thermal equilibrium with the lattice (i.e. to the LS state),
hinders the efficiency of light irradiation. The incomplete
conversion achieved at 4 K (98 $\%$) may also be due to the
partial penetration of the light throughout the sample volume. It
takes more than 8 hours to reach a conversion efficiency around 22
$\%$ (Fig. \ref{Fig:eff}).
The efficiency seems to scale with $T_{irr}$ following a power-law as shown in Fig. \ref{Fig:eff}, suggesting that light penetration in the solid may be a key issue in determining the photoconversion time.


The decay time of the light-induced HS state, driven by phonon
processes, was derived at three different temperatures and found
to decrease with increasing temperature. In order to compare with
previous results in the literature we considered just the data
above 30 K and fit them with a thermally activated law
\cite{Beni2006,Li2008}
\begin{equation}
T_{dec}(T)=T_{0}\cdot e^{\frac{\Delta E}{k_{B}T}} \,\, .
\label{arrhenius}
\end{equation}
Even if characterized by a large uncertainty, one can roughly
estimate an activation barrier $\Delta E\simeq 350$ K for T $>$ 30
K (data are reported in the Supplementary Material), which is
consistent with the 390 K activation barrier previously reported
\cite{Dapporto2008}. Some computational studies based on the
Density Functional Theory  have been carried out in Co-based
valence tautomers
\cite{Adams1997,Bencini2003,Sato2010,Starikov2011} similar to the
one used in this work in order to estimate the barrier between the
LS and HS states and a reasonable agreement with the experimental
values was found.

The information on the temperature evolution of the VT spin state
and on the HS lifetime derived by SQUID magnetometry can be
suitably complemented by that derived from local probes, as nuclei
and muons, which can provide information on the spin dynamics of
HS molecules and on how it is affected by external stimuli. $^1$H
NMR spin-lattice relaxation is nearly constant above 200 K as it
is expected for weakly coupled spins in the high temperature limit
\cite{slichter}. The slight magnetic field dependence of
$1/T_{1H}$ observed at high temperatures (Fig.\ref{Fig:T11H}),
suggests the presence of low-frequency diffusive spin
fluctuations, similar to those of low-dimensional paramagnets
\cite{Carretta2004,FBranzoli2009}, or due to molecular motions.
Below 200 K $1/T_{1H}$ drops, due to the VT transition. At 100 K,
when most VT molecules are diamagnetic, $1/T_{1H}$ is reduced by
50 \%, since the mean squared amplitude of the fluctuating local
fields at the $^1$H nuclei is strongly diminished by the
conversion.


At low temperatures (5 - 80 K) $^1$H $1/T_{1H}$ displays a peak
around 20 K, suggesting a slowing down of the spin fluctuations to
frequencies approaching the nuclear Larmor frequency for that
fraction of VT molecules which are still in the HS state at low
temperature, those yielding the non-zero contribution to
$\chi_{M}\cdot T$ =0.24 emu$\cdot$K/mol. This behaviour is similar
to that observed in molecular magnets, where a peak in $1/T_1$ is
detected at temperatures of the order of the energy separation
between the ground and the first excited states
\cite{FBranzoli2009,Branzoli2009,Papinutto2005} (for a complete list see Ref. \cite{MRbook}, pg 29-70, and references therein). The magnetic
field dependence shown in Fig. \ref{Fig:peakfield} confirms this
hypothesis, since the peak intensity decreases with the nuclear
Larmor frequency(see also Eq. \ref{BPP} in the following).

Thus, the fraction of VT in the HS state at a given
temperature can be described as an ensemble of paramagnetic
impurities embedded in a diamagnetic matrix, whose spin
progressively slow down to frequencies approaching the tens of MHz
range around 20 K. If one considers that the thermal population of
the HS molecule states is given by $p(T)=exp(-E_{A}/T)/Z,$ with
$E_{A}$ an effective barrier between the low-energy levels
involved in the HS fluctuations and $Z$ the corresponding
partition function, one can weight the amplitude of the local
field fluctuations driving the spin-lattice relaxation by $p(T)$.
Then, by assuming a Lorentzian broadening of the HS state due to
phonon scattering processes, one can write for the spin-lattice
relaxation rate \cite{bppBPP1948,Corti1997,Carretta1993}:
\begin{center}
$\frac{1}{T_{1}}\cong\frac{\gamma^{2}}{2}\frac{\delta
h_{\bot}^{2}}{\omega_{n}\Delta} p(T) T \biggr[
\arctan\left[\omega_{n}\tau_{0}\exp\left(\frac{E_{A}+\Delta}{T}\right)\right]+$
\begin{equation}
\label{BPP}
-\arctan\left[\omega_{n}\tau_{0}\exp\left(\frac{E_{A}-\Delta}{T}\right)\right]\biggl].
\end{equation}
\end{center}
We have considered a rectangular distribution for the activation
energy $E_A$ with width $\Delta$. $\tau_{0}$ is the high
temperature limit of the correlation time, $\delta h_{\bot}^{2}$
is the mean squared amplitude  of the local field fluctuations,
$\gamma$ the nuclear gyromagnetic ratio and $\omega_n$ the nuclear
Larmor frequency. The fit yields an average barrier $E_A\simeq 90$
K and a smaller $\Delta\simeq 60$ K (see Table \ref{Tbl:BPPCo}),
as one may deduce from the narrow peak reported in
Fig.\ref{Fig:T11H}. Notice that $E_A$, significantly smaller than
$\Delta E$, is not a barrier separating the LS and HS states. As
it can be noticed from Fig. \ref{Fig:peakfield} light irradiation
slightly enhances the proton spin-lattice relaxation rate without
affecting the temperature dependence close to the peak.



$\mu$SR measurements, both in ZF and LF, show that muon relaxation
rate is not much sensitive to the VT transition (Fig.
\ref{Fig:lambdazf}), suggesting that around 160 K the spin
dynamics are much faster than the muon lifetime ($\simeq
2.2\mu$s). On the other hand, at low temperature, around 20 K,  ZF
and LF relaxation rates show a neat peak (Fig. \ref{Fig:lambdazf})
analogous to the one detected by $^1$H spin-lattice relaxation
rate. A similar behavior of the muon spin relaxation rates has
been also reported for spin crossover systems \cite{Blundell2004}
as well as for other systems characterized by a spin freezing
\cite{Campbell1994,Murnick1976}. In the LF measurements, where the
nuclear dipolar contribution to the muon relaxation is suppressed,
one observes a temperature dependence which is rather similar to
that of $1/T_{1H}$. In fact, by fitting $\lambda_{LF}$ data with
Eq. \ref{BPP}, one finds values for $E_A= 77\pm 5$ K and $\Delta=
54\pm 4$ K, which are close to those derived from $1/T_{1H}$,
confirming the hypothesis of the spin freezing of the HS
molecules.

To get a deeper insight on the HS molecules spin dynamics we now
turn to the discussion of $^{59}$Co spin-lattice relaxation rate
which is a more sensitive to spin fluctuations, thanks to the
hyperfine coupling with the paramagnetic ion. As pointed out in
Sect. II.2, the signal being detected is the one from $^{59}$Co
nuclei belonging to VT in the LS state. In fact, as it is shown in
Fig. \ref{Fig:CoSignal} the normalized signal intensity decreases
above the LS-HS transition temperature and vanishes at high
temperatures.

It is pointed out that at 180 K $1/T_{1Co}$ is much larger than
$1/T_{1H}$ (Fig. \ref{Fig:T1CoSpectrum}), which is unexpected
since the relaxation rate should scale with the square of the
gyromagnetic ratio (4.21 times larger for $^1$H) and the average
hyperfine coupling is expected to be larger for $^1$H rather than
for $^{59}$Co nuclei in the LS state, which on average are farther
from the centre of the molecule. Hence, it is likely that the
relaxation rate of $^{59}$Co nuclei in LS VT is enhanced through
nuclear spin diffusion by the relaxation of $^{59}$Co nuclei in HS
VT. In other terms, $1/T_{1Co}$ is probing a weighted average of
the relaxation of the nuclei in LS and HS molecules but since the
relaxation in the latter ones is much faster one is basically
probing the relaxation of $^{59}$Co nuclei in LS molecules.

The spin transition driven by the electron transfer causes a
marked reduction of $^{59}$Co nuclear spin-lattice relaxation
rate, with a decrease of $1/T_{1Co}$ by more than two orders of
magnitude between 180  and 50 K. At lower temperature, around 25
K, a small peak is observed (see Fig. \ref{Fig:T1Co}). This peak
is analogous to the one observed for $1/T_{1H}$. The points in the
range (5 - 55 K) were fit using equation \ref{BPP} (see Table
\ref{Tbl:BPPCo}) and a value for the $E_A$ barrier similar to that
derived from $^{1}$H or muon spin relaxation was derived.

\begin{table}
\begin{tabular}{|c|c|c|c|c|}
\hline
  &$\delta h_{\bot}^{2} (T^{2})$ &$\tau_{0} (s) $ &$E_A (K) $ &$\Delta (K)$\tabularnewline
\hline
 $^{59}$Co OFF  & $4.1 \pm 0.9 \cdot 10^{-7}$ &  $1.7 \pm 0.1 \cdot 10^{-10}$ &    $53 \pm 16$& $43 \pm17$                  \tabularnewline
\hline
$^{59}$Co ON &  $1.8 \pm 0.3 \cdot 10^{-6}$&   $1.1 \pm 0.5 \cdot 10^{-10}$ &    $74\pm 17$&     $60 \pm 17$             \tabularnewline
\hline
 $^{59}$Co DIF. &    $1.4 \pm 0.3 \cdot 10^{-6}$&     $5.6 \pm 3.8 \cdot 10^{-11}$&  $88\pm 20$&   $72\pm 18$               \tabularnewline
\hline
$^1$H OFF &     $2.8 \pm 0.4 \cdot 10^{-6}$&  $3.5 \pm 0.8 \cdot 10^{-11}$&   $90 \pm 12$&    $60 \pm 10$                \tabularnewline
\hline
LF $\mu$SR &      $3.2 \pm 0.1 \cdot 10^{-10}$&   $1.5 \pm 0.9 \cdot 10^{-10}$&    $77 \pm 5$&     $54 \pm 4$               \tabularnewline
\hline
\end{tabular}
\caption{Fit results according to Eq. \ref{BPP} for $^{59}$Co, $^{1}$H and muon relaxation data.} \label{Tbl:BPPCo}
\end{table}


Infrared light irradiation enhances the $1/T_{1Co}$ at
low temperature (Fig. \ref{Fig:T1Co}) because of
the increase in the number of molecules in the HS state, a much
more marked effect than that obtained on protons. Moreover, under
light irradiation $1/T_{1Co}$ value remains significantly lower
than at $T\simeq 180$ K, pointing out that the conversion
efficiency in the NMR experiments is lower than in SQUID
magnetization measurements. This could be due to the limited
penetration of the infrared radiation over the whole volume of the
sample used in the NMR experiments. Eventhough, the enhancement of
$1/T_{1Co}$ allows to probe the spin dynamics of those molecules
which have been promoted to the HS state.

The low-temperature behaviour of $1/T_{1Co}$ upon irradiation can
still be fit with Eq. \ref{BPP} below 50 K with similar values for
the activation energies (see Tab.\ref{Tbl:BPPCo}). In order to
consider the spin dynamics due to the light-induced HS molecules
only, we have subtracted the behaviour of $1/T_{1Co}$ before
irradiation from the one in presence of irradiation. The resulting
temperature dependence, shown in the inset of Fig. \ref{Fig:T1Co},
was fit again with Eq. \ref{BPP} and the best fit was achieved for
$E_A= 88\pm 20$ K and $\Delta= 72\pm 18$ K. Thus, it is noticed
that there is a low-frequency (MHz range) dynamic which persists
at low temperature with a barrier which is similar to that derived
from $^1$H spin-lattice relaxation rate measurements. These values
are significantly lower than the barrier $\Delta E$ estimated from
the decay of the magnetization after light irradiation with the
SQUID magnetometer. Indeed while this latter is the barrier needed
to switch between the two VT states, the spin fluctuations probed
by NMR and $\mu$SR relaxations may involve transitions among
states of the HS molecules characterized by a different
energy separation. In view of future application
these low-frequency fluctuations must be somewhat reduced or even
removed by playing with the chemical structure of the molecule.

Finally we discuss the effect of an external pressure on the VT
conversion. In Fig. \ref{Fig:Mag_P} one observes that moderate
pressures, of the order of a few kbar, shift the transition
detected by $M(T)$ measurements close to room temperature. The
$1/T_{1H}$ measurements performed at 300 K do support the
magnetization results. As it is shown in Fig. \ref{Fig:1H_P},
$^1$H spin-lattice relaxation rate decreases with increasing
pressure, indicating a progressive transition of the VT from the
HS to the LS state, ending at about 10 kbar, while for higher
pressures $1/T_{1H}$ does not change and the system remains in the
LS state. In fact, the LS state is favoured by high pressures
since it is characterized by a smaller volume giving rise to
larger crystal field splittings with respect to the HS state
\cite{Tezgerevska2014,Evangelio2005,Dei2010b}. The observation of
a pressure induced transition at room temperature  is relevant for
the application of VT as sensors.

\section{Conclusion}

By means of macroscopic magnetization and of nuclear and muon
spin-lattice relaxation rate measurements we have investigated the magnetic properties of a cobalt compound
showing Valence Tautomerism. In particular we have focused on the effect of infrared light
irradiation at low temperatures and that of an external
pressure on the interconversion between the two spin
states associated with different intramolecular charge
distribution. The SQUID magnetometry data result in good agreement with earlier reports and  and point out two important
facts: quite long build-up times (up to 9 hours) are needed to
reach a conversion efficiency close to 100 \% and the decay rate
of light-induced magnetization is determined by an energy barrier
which leads to many hours lifetimes for the HS state at low
temperatures. On the other hand, by means of NMR and $\mu$SR we
have evidenced that spin fluctuations in the MHz range persist at
low temperature. By increasing the number of HS molecules with
infrared light irradiation the temperature evolution of the spin
dynamics does not change significantly. Hence, although the
light-induced HS state of VT can persist for hours at $T\simeq 10$
K, spin dynamics can be faster approaching timescale of seconds at temperatures around few Kelvin degrees. These results
show that application of this material as a logic unit still needs further developments. On the other hand, we have
shown that the VT transition can suitably be tuned by an external
pressure and reaches room temperature at $P\simeq 10$ kbar, an
observation which could pave the way to immediate application of
valence tautomers as sensors. The marked increase in the crossover
temperature with pressure suggests that the spin dynamics
could also be significantly affected and possibly reduced by an
external pressure.

\section*{Acknowledgements}
Marco Moscardini is gratefully acknowledged for his technical
assistance during the setup of the light irradiation experiments
in the SQUID magnetometer.
James Lord is gratefully acknowledged for his assistance during $\mu$SR
experiments carried out at the ISIS facility.
Lorenzo Sorace is widely acknowledged for the fruitfull discussion during the revision of the manuscript.


\end{document}